\documentclass[twocolumn,showpacs,amsmath,amssymb,prl]{revtex4}
\usepackage{graphicx}

\begin{document}
\title{Interface thermal resistance between dissimilar anharmonic lattices }

\author{Baowen Li$^{1,2,4}$}
\email{phylibw@nus.edu.sg} \author{Jinghua Lan$^1$} \author{Lei
Wang$^{3,4}$}

 \affiliation{$^1$ Department of
Physics, National University of Singapore, Singapore 117542,
Republic of
Singapore\email{phylibw@nus.edu.sg}\\
$^2$NUS Graduate School for Integrative Sciences and Engineering,
Singapore 117597, Republic of Singapore
\\
$^3$ Temasek Laboratories, National University of Singapore,
Singapore 119260, Republic of Singapore\\
$^4$ The Beijing-Hong Kong-Singapore Joint Center for Nonlinear
and Complex Systems (Singapore), Singapore 117542}
\date{Published in Phys. Rev. Lett, {\bf 95}, 104302 (2005)}

\begin{abstract}
We study interface thermal resistance (ITR) in a system consisting
of two dissimilar anharmonic lattices exemplified by
Fermi-Pasta-Ulam (FPU)
 model and Frenkel-Kontorova (FK) model.  It is found that the ITR is asymmetric, namely,
it depends on how the temperature gradient is applied. The
dependence of the ITR on the coupling constant, temperature,
temperature difference, and system size are studied. Possible
applications in nanoscale heat management and control are
discussed.
\end{abstract}
\maketitle

When heat flows across an interface between two different
materials (phases), there exists a temperature jump at the
interface, from which we can define an interface thermal
resistance (ITR):
\begin{equation}
R\equiv\frac{\Delta T}{J}, \label{eq:Rheat}
\end{equation}
where $J$ is the heat flux density, namely, the heat flow across a
unit area in unit time, $\Delta T$  is the temperature difference
between two sides of the interface. This problem has been caught
attention as early as 1941 when Kapitza\cite{Kapitza41} discovered
the temperature jump at an interface between solid and liquid.
Continuous efforts have been devoted to this problem (see Review
article \cite{Pohl89}, and the references therein) since then.
More recently, the temperature jump between liquid and vapor has
also been observed experimentally\cite{Fang} and studied by
computer simulations\cite{Vapour}.

A general theory to describe the temperature jump at interface of
different materials (in different phases) is still lacking. There
are two popular theories for harmonic lattice: diffuse mismatch
theory\cite{Pohl89} and acoustic mismatch theory\cite{Little59}.
The acoustic mismatch theory regards the two media as two elastic
continua, while the diffuse mismatch theory assumes that at the
interface all phonons are diffusively scattered. When anharmonic
interaction is taken into account, superposition theorem fails,
thus no analytic theory can be worked out. However, as we shall
show in the following that, if the \emph{anharmonic} interaction
is considered, more interesting phenomena arise, the problem
becomes theoretically more challenging, and of course it is closer
to reality because harmonic is just a first order approximation.
On the other hand, as the rapid development of nanotechnology, low
dimensional nano scale systems such as nanowires and nanotubes can
be easily fabricated in the lab. At the nano scale, the systems
are of finite size, more precisely, they are discrete, and
therefore, the continuous theory such as the acoustic mismatch
theory will definitely not be suitable for such systems. The
interface resistance in such nano systems becomes more and more
important and has potential applications in nanoscale heat control
and management\cite{ReviewNano}.

In this Letter, we study the ITR in the system comprising two
dissimilar \emph{anharmonic} lattices. The system, illustrated in
Fig. 1, consists of two chains of $N$ oscillators. The left part
is a chain of $N$ harmonic oscillators on a substrate whose
interaction is represented by a sinusoidal on-site potential. The
right part is a chain of $N$ anharmonic oscillators. The two parts
are connected by a spring of constant $k_{int}$, and the
Hamiltonian of the system is,
\begin{equation}
H=H_{FK}+H_{FPU}+\frac{1}{2}k_{int}(x_{N+1}-x_{N}-a)^2,
\label{Hsys}
\end{equation}
where $H_{FK}$ is the Hamiltonian of the left part which is in
fact the Frenkel Kontorova (FK) model, $
H_{FK} =\sum_{i=1}^{N}
\frac{p^2_i}{2m}+\frac{1}{2}k_{FK}(x_i-x_{i-1}-b)^2-\frac{V}{(2\pi)^2}\cos
2\pi x_i. $
$H_{FPU}$ is the Hamiltonian of the right part which is the
Fermi-Pasta-Ulam (FPU) model, $
H_{FPU}=\sum_{i=N+1}^{2N}\frac{p^2_i}{2m}+k_{FPU}\left[\frac{1}{2}(x_{i+1}-x_i-a)^2+
\frac{\beta}{4}(x_{i+1}-x_i-a)^4\right]. $
Fixed boundaries are used, i.e. $x_0=0$ and $x_{2N+1}=Nb+(N+1)a$.
The FPU model is a representative anharmonic lattice without
on-site potential and the FK model is the one with on-site
potential. Both models have been widely used to study different
problems in condensed matter physics and nonlinear
dynamics\cite{ChaosFocus,Braun}. In particular, the FPU model has
played an important role in the development of computational
physics and nonlinear dynamics\cite{ChaosFocus}. Recent years have
witnessed an increasing interests in the study of heat conduction
with these two models\cite{FPU,HLZ00,FK,Review,TPC02}.

\begin{figure}
\includegraphics[width=\columnwidth]{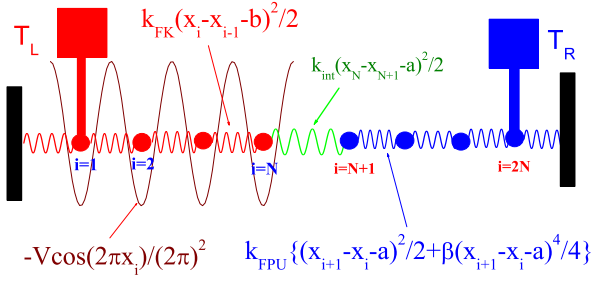}
\vspace{-.8cm} \caption{Configuration of the system. The left part
is a Frenkel-Kontorova model and the right one is a
Fermi-Pasta-Ulam model. The two parts are connected by a spring of
constant $k_{int}$. The two particles on the left and right ends
are put into contact with heat baths of temperature $T_L$ and
$T_R$, respectively.} \label{Fig:config}
\end{figure}

In our simulation, Nos\'e-Hoover heat baths are used. The system
parameters are
 $k_{FK}=1$, $\beta=1$, $a=b=1$,
$m=1$, and $V=5$. $T_L=T_0(1+\Delta), T_R=T_0(1-\Delta)$.
$k_{FPU}=0.2$ unless otherwise stated. The local temperature at
site $n$ is defined as $T_n = m \langle \dot{x_n}^2 \rangle $,
where $\langle \hspace{0.2cm} \rangle$ stands for temporal
average. We calculate the physical quantities after a time that is
long enough to allow the system to reach a non-equilibrium steady
state where the local heat flux is constant along the chain.

\begin{figure}
\includegraphics[width=\columnwidth]{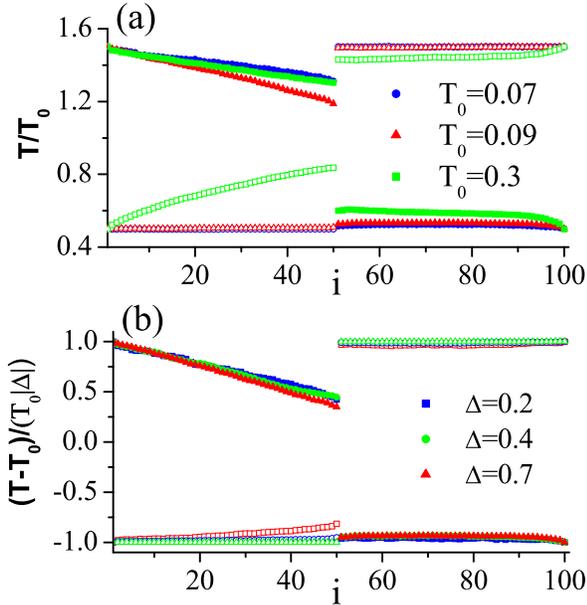}
\vspace{-1.cm} \caption{\label{fig:Temperature}(a) $T/T_0$ versus
lattice site for $T_0=0.07, 0.09$, and $0.3$. $|\Delta|=0.5$.  (b)
$(T-T_0)/(T_0|\Delta|)$ versus lattice site, for different
$|\Delta|=0.2, 0.4,$ and $0.7$ with fixed $T_0=0.09$. The solid
symbols are for the cases of $\Delta>0$, and the empty ones are
for the cases of $\Delta<0$.  In both (a) and (b) $N=50$.}
\end{figure}

Fig.\ref{fig:Temperature}(a) shows the normalized temperature,
$T/T_0$, along the lattice site for three different temperatures,
$T_0$=0.07, 0.09, and 0.3, with fixed $|\Delta|= 0.5$. Fig.
\ref{fig:Temperature}(b) shows $(T-T_0)/(T_0|\Delta|)$ versus
lattice site for $|\Delta|$=0.2, 0.4, and 0.7 with fixed
$T_0=0.09$. It is obvious that, in all cases, there exists a
temperature jump (discontinuity) at the interface. The jump
depends not only on temperature $T_0$ but also the temperature
difference $\Delta$. The most interesting thing is that,
\textit{the jump is asymmetric, namely, it depends on whether
$\Delta$ is positive or negative.}

\begin{figure}
\includegraphics[width=\columnwidth]{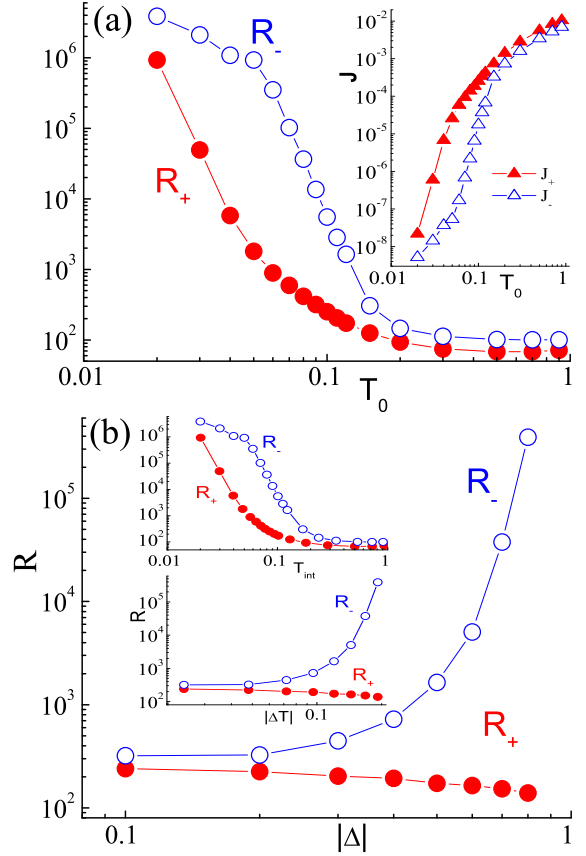}
\vspace{-1.cm} \caption{(a) $R_{\pm}$ versus $T_0$.
$|\Delta|=0.5$. Inset is heat current $J_{\pm}$ versus $T_0$. (b)
$R_{\pm}$ versus $|\Delta|$ for $T_0=0.12$. Inset of (b) is
$R_{\pm}$ versus interface temperature
$T_{int}=(T_{int}^L+T_{int}^R)/2$, and $R_{\pm}$ versus interface
temperature jump, $\Delta T$. In all cases $N=50$.
}\label{fig:R-profile}
\end{figure}

The asymmetric feature is well captured by ITR shown in Fig.
\ref{fig:R-profile}.  $R_{\pm}$ denotes the ITR when $\Delta>0$
and $\Delta<0$, respectively. Fig\ref{fig:R-profile}(a) shows
that, at intermediate temperature, $T_0\in(0.02,0.1)$, the ratio,
$R_-/R_+$, changes from about 10 to 1000. Both $R_+$ and $R_-$
decrease with $T_0$ until $T_0\approx 0.2$ and then become
approximately constants. The dependence of $R_{\pm}$ can be
explained from the temperature dependence of the phonon spectra,
as we shall discuss fully later on. In the inset of
Fig.\ref{fig:R-profile}(a), we show the heat currents $J_{\pm}$
versus temperature $T_0$. $J_{\pm}$ is the heat flux for $\Delta
>0$ and $\Delta<0$, respectively. Heat currents increase very fast
(in several orders of magnitude) as $T_0$ is increased over two
orders of magnitude.

In Fig. \ref{fig:R-profile}(b), we draw $R_{\pm}$ versus
$|\Delta|$ for $T_0=0.12$. It shows that as $|\Delta|$ increases,
$R_+$ does not change too much, it is always in the order of 200,
whereas $R_-$ increases more than three orders of magnitude. In
the case of $|\Delta|=0.8$, $R_-/R_+$ can be larger than 1,000 and
$|J_+| \approx 2,000 |J_-|$. The dependence of ITR on the
interface temperature, $T_{int}$ and the interface temperature
jump, $\Delta T\equiv T^L_{int}-T^R_{int}$ are shown in the inset
of Fig. \ref{fig:R-profile}(b).  Where $T_{int}^{L,R}$ is the
temperature for the particle on the left/right side of the
interface, respectively.

An obvious conclusion from above results is that the ITR between
dissimilar \emph{anharmonic} lattices is {\it asymmetric}.

To understand the physical mechanism of this phenomenon, we need
to invoke the energy band theory. However, due to the presence of
anharmonicity, an analytic approach seems impossible. We shall
rather take a qualitative approach, which can also give us useful
information in two extreme cases, namely, low temperature limit
(regime) and high temperature limit (regime).

{\bf FK model}.  At very low temperature, the particles are
confined in the valleys of the on-site potential. By linearizing
equations of motion one can easily obtain the phonon band, $
\sqrt{V}<\omega^L_{FK} <\sqrt{V+4k_{FK}}.$ On the other hand, in
the high temperature limit, the particles have large enough
kinetic energies to jump out the valleys. The on-site potential
becomes negligible, the FK model degenerates to an harmonic one,
$0<\omega^H_{FK}<2\sqrt{k_{FK}}$.

{\bf FPU model}. There exists a threshold temperature
$T_{th}(\approx 0.1) $\cite{HLZ00}, below which the FPU model
becomes a harmonic one,  one thus has
$0<\omega^L_{FPU}<2\sqrt{k_{FPU}}$. When $T_0>>T_{th}$, the
anharmonic term is dominant. In this regime, a rough theoretical
estimate yields $0<\omega^H_{FPU}<C_0 (Tk_{FPU}\beta)^{1/4}$ with
$C_0=2\sqrt{2\pi}\Gamma (3/4)3^{1/4}/\Gamma(1/4)\approx 2.23$,
where $\Gamma$ is the Gamma function.

\begin{figure}
\includegraphics[width=\columnwidth]{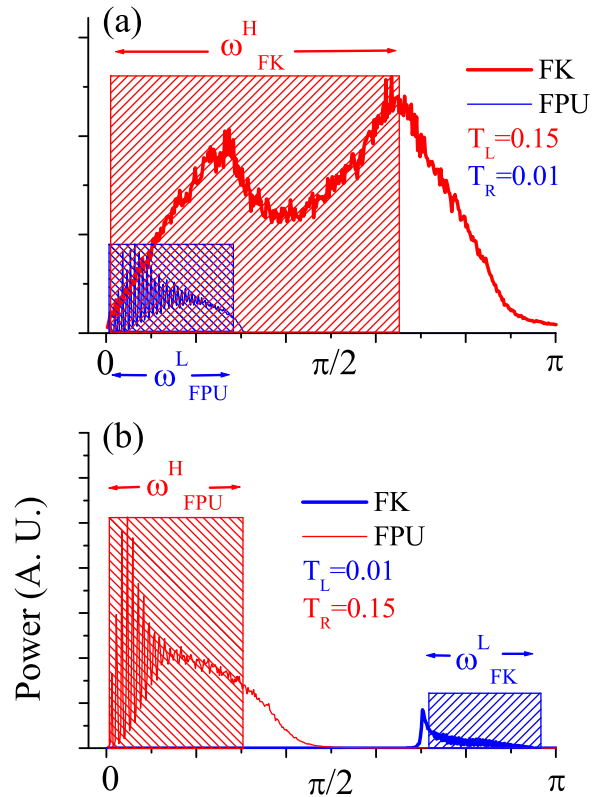}
\vspace{-1.2cm} \caption{Phonon spectra of the two particles at
interface and schematic phonon bands for the FK and the FPU models
in two extremes. (a) $T_L=0.15$ and $T_R=0.01$.  (b) $T_L=0.01$
and $T_R=0.15$. The shadow regions are the analytical estimates
for the FK and FPU models. (see text for more explanation.) }
\label{fig:spectra}
\end{figure}

The spectra of the interface particles are shown in
Fig.\ref{fig:spectra}(a) and (b) for $\Delta>0$ and $\Delta<0$,
respectively, and compared with the above analytical analysis (the
shadow regions). In the first case, when $T_L=0.15$ and
$T_R=0.01$, we can approximately regard the FK lattice as at high
temperature limit, and the FPU lattice at low temperature regime.
With $k_{FK}=1$ and $k_{FPU}=0.2$, we
have, $\omega^H_{FK} \in[0, 2]$, and $\omega^L_{FPU} \in[0,
0.89]$, which are quite close to the numerical ones.

On the other hand, when the temperatures of the two thermal baths
are swapped. Approximately, the FK lattice is at low temperature
limit, and the FPU is at high temperature regime. In this case,
according to above analysis, the phonon band of the FK model is
$\omega^L_{FK}\in[2.24, 3.00]$ for $V=5$ and $k_{FK}=1$, and
$\omega^H_{FPU} \in[0, 0.94]$ with $T\approx 0.15$ and $\beta=1$
for the FPU model. They are also very close to the numerical
results shown in Fig.\ref{fig:spectra}(b).

It is clear from Fig.\ref{fig:spectra} that, in the case of
$T_L>T_R$, the phonon band of the FK lattice overlaps that one of
the FPU lattice, therefore, heat can easily flow from the FK
lattice to the FPU lattice as is demonstrated in
Fig.\ref{fig:R-profile} by small $R_+$ and large $J_+$.
Conversely, when $T_L<T_R$, by appropriately chosen parameters, a
large gap between the phonon bands of the FK lattice and the FPU
lattice can be formed (see Fig.\ref{fig:spectra}(b)), which
inhibits heat flow from the FPU lattice to the FK lattice as is
manifested by a large $R_-$ and small $J_-$ in
Fig.\ref{fig:R-profile}. However, as $T_0$ increases, the gap
becomes narrower and narrower, and eventually disappears when
$T_0$ surpasses a certain value. Indeed, when $T_0$ is large
enough the two phonon bands overlap leading to a constant $R_-$ as
is seen in Fig.\ref{fig:R-profile}(a).

\begin{figure}
\includegraphics[width=\columnwidth]{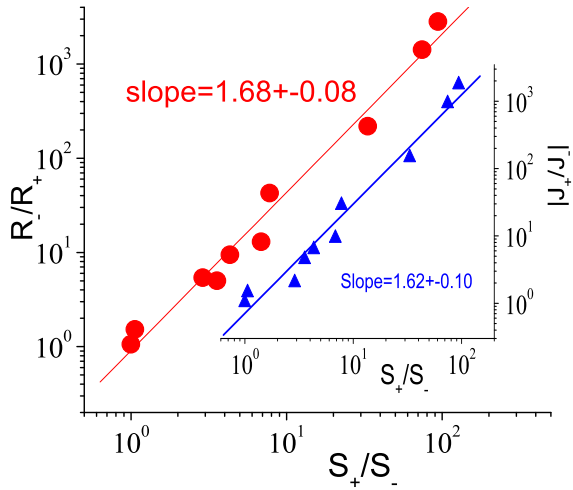}
\vspace{-1.2cm} \caption{$R_-/R_+$ versus $S_+/S_-$. Inset,
$|J_+/J_-|$ versus $S_+/S_-$. } \label{Fig:RJS}
\end{figure}

In  order to quantify above asymmetric ITR, heat currents, and
find the relationships with the overlap of the phonon bands of the
two lattices, we introduce,
\begin{equation}
S_{\pm}  =\frac{\int_0^{\infty}
P_l(\omega)P_r(\omega)d\omega}{\int_0^{\infty}P_l(\omega)d\omega
\int_0^{\infty}P_r(\omega)d\omega}. \label{eq:Spm}
\end{equation}
$S_{\pm}$ correspondes to the overlap for $\Delta>0$ and
$\Delta<0$, respectively. Note that
 $\int_0^{\infty}P_{l,r}d\omega=T^{L,R}_{int}$. In
Fig.\ref{Fig:RJS} we plot $R_-/R_+$ versus $S_+/S_-$ and in the
inset we show $|J_+/J_-|$ versus $S_+/S_-$. Best fit gives
$R_-/R_+ \sim \left(S_+/S_-\right)^{\delta_R}$ with
$\delta_R=1.68\pm 0.08$, and $|J_+/J_-| \sim
\left(S_+/S_-\right)^{\delta_J}$, with $\delta_J=1.62\pm 0.10$.
This picture indeed illustrates that the ITR and heat current
correlate strongly with the overlap of the phonon spectra of the
two interface particles.

\begin{figure}
\includegraphics[width=\columnwidth]{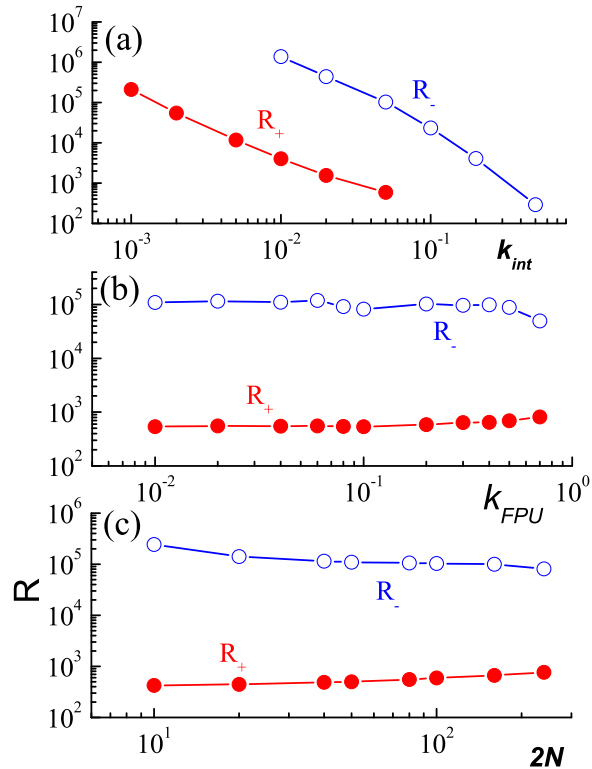}
\vspace{-1.2cm} \caption{\label{fig:Rparam}(a) $R_{\pm}$ versus
the coupling constant $k_{int}$. (b) $R_{\pm}$ versus $k_{FPU}$.
(c)$R_{\pm}$ versus the system size, $N$. For all three cases,
$T_0=0.07$ and $|\Delta|=0.5$.  $N=50$ for (a) and (b).
$k_{int}=0.05$ for (b) and (c).}
\end{figure}

As the system contains many adjustable parameters, it is worth
investigating the dependence of the ITR on these parameters. Fig.
\ref{fig:Rparam}(a) is $R_{\pm}$ versus $k_{int}$, which shows
both $R_{\pm}$ decreases with increasing $k_{int}$. This means
that strong coupling favorites heat transport. Fig
\ref{fig:Rparam}(b) shows $R_{\pm}$ versus $k_{FPU}$ and
Fig.\ref{fig:Rparam}(c) demonstrates the finite size effect. $R_+$
increases slightly whereas $R_-$ decreases slightly with $N$. This
can be understood from the fact that with fixed $T_+$ and $T_-$,
when $\Delta>0$, the larger $N$ the smaller the $T^L_{int}$, thus
the larger $R_+$; conversely, when $\Delta<0$, the larger $N$, the
larger $T^L_{int}$ thus the smaller $R_-$.

The asymmetry of the ITR and the heat current in the model studied
is due to the anharmonicity (nonlinearity) of the two lattices. In
the low temperature limit, such as $T_0<0.01$, in which both the
FPU  and the FK models can be approximated by harmonic lattices,
the asymmetry property will vanish because the phonon band of
harmonic lattice is temperature independent. On the other hand, in
the high temperature regime, in particular, when $T_0>>V/(2\pi)^2$
such that the on-site potential can be neglected, then the phonon
band of the FK  model becomes temperature independent. In this
case, the asymmetry effect becomes minimal as we can see from Fig.
3 when $T_0>0.2$. However, the asymmetry will never disappear as
long as the anharmonic term in the FPU model is still present.

In summary, we have studied the ITR between two dissimilar
\emph{anharmonic} lattices. It is found that the ITR is {\it
asymmetric}, which is believed to be very general as the two {\it
anharmonic} lattices used in this Letter are two representative
ones widely studied in different fields of physics. The asymmetric
property might be useful in heat control and
management\cite{ReviewNano}. In particular, the very high ITR
might find applications in building thermal insulator.

BL is supported in part by a FRG of NUS and the DSTA under Project
Agreement POD0410553. LW is supported by DSTA  POD0001821.


\begin{thebibliography}{01}


\bibitem{Kapitza41} P. L. Kapitza, J. Phys. USSR {\bf 4}, 181 (1941).

\bibitem{Pohl89}E. T. Swartz and R. O. Pohl, Rev. Mod. Phys. {\bf 61}, 605 (1989).

\bibitem{Fang}G. Fang and C. Ward, Phys. Rev. E \textbf{59}, 417,
441 (1999).

\bibitem{Vapour} A. Rosjorde et al. J. Colloid. Interface Sci. {\bf 232}, 178 (2000); {\bf 240},
355 (2001); J. M. Simon et al. J. Phys. Chem. B {\bf 108}, 7186
(2004).


\bibitem{Little59} W. A. Little, Can. J. Phys. {\bf 37}, 334 (1959).

\bibitem{ReviewNano}D. G. Cahill et al. J. App. Phys. {\bf 93}, 793 (2003).

\bibitem{ChaosFocus} D. K. Campbell, P. Rosenau, and G. Zaslavsky ed.
"Focus Issue on 50'th anniversary of Fermi-Pasta-Ualam
model", Chaos {\bf 15}, no. 1, 015101-015121 (2005).

\bibitem{Braun} O. M. Braun and Yu. S. Kivshar, ``{\it The Frenkel-Kontorova Model: Concepts, Methods,
and Applications}, Springer-Verlag, Berlin, 2003.

\bibitem{FPU} H. Kaburaki and M. Machida, Phys. Lett. A\textbf{181}, 85 (1993); S. Lepri \textsl{et al},
Phys. Rev. Lett. \textbf{78}, 1896 (1997); A. Fillipov, B. Hu, B.
Li, and A. Zeltser, J. Phys. A \textbf{31}. 7719 (1998), S. Lepri,
Phys. Rev. E \textbf{58}, 7165 (1998); K. Aoki and D. Kusnezov,
Phys. Rev. Lett \textbf{86}, 4029 (2001); A. Pereverzev, Phys.
Rev. E, \textbf{68}, 056124 (2003).

\bibitem{HLZ00}B. Hu, B. Li, and H. Zhao, Phys. Rev. E
\textbf{61}, 3828 (2000).

\bibitem{FK} B. Hu, B. Li, and H. Zhao, Phys. Rev. E \textbf{57}, 2992
(1998); A. V. Savin and O. V. Gendelman, Phys. Rev. E \textbf{67},
041205 (2003).

\bibitem{Review}F. Bonetto \textsl{et al.},
in ``Mathematical Physics 2000,'' A. Fokas \textsl{et al.} (eds)
(Imperial College Press, London, 2000) (pp. 128-150); S. Lepri
\textsl{et al.}, Phys Rep. \textbf{377}, 1 (2003).

\bibitem{TPC02}M. Terraneo, M. Peyrard, and G. Casati, Phys. Rev. Lett \textbf{88}, 094302 (2002).
B. Li, L. Wang, and G. Casati, Phys. Rev. Lett, \textbf{93},
184301 (2004).

\end{thebibliography}
\end{document}